\documentclass[twoside,a4paper,11pt]{sea9}
\usepackage{graphicx}
\begin{document}
\pagenumbering{arabic}
\pagestyle{myheadings}
\thispagestyle{empty}
{\flushleft\includegraphics[width=\textwidth,viewport=58 650 590 680]{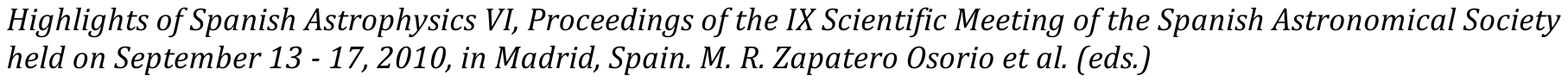}}
\vspace*{0.2cm}
\begin{flushleft}
{\bf {\LARGE
An X-ray Study of Local Infrared Bright Galaxies
}\\
\vspace*{1cm}
Miguel Pereira-Santaella$^{1}$,
Almudena Alonso-Herrero$^{1,2}$,
Elena Jim\'enez-Bail\'on$^{3}$,
George H. Rieke$^{4}$,
Mar\'ia Santos-Lleo$^{5}$,
and
Martin Ward$^{6}$
}\\
\vspace*{0.5cm}
$^{1}$
Departamento de Astrof\'isica, Centro de Astrobiolog\'ia, CSIC/INTA, Carretera de Torrej\'on a Ajalvir, km 4, 28850, Torrej\'on de Ardoz, Madrid, Spain\\
\texttt{pereira@damir.iem.csic.es}\\
$^{2}$
Associate Astronomer, Steward Observatory, University of Arizona, 933 North Cherry Avenue, Tucson, AZ 85721, USA \\
$^{3}$
Instituto de Astronom\'ia, Universidad Nacional Aut\'onoma de M\'exico, Apartado Postal 70-264, 04510 Mexico DF, Mexico\\
$^{4}$
Steward Observatory, University of Arizona, 933 North Cherry Avenue, Tucson, AZ 85721, USA\\
$^{5}$
XMM-Newton Science Operations Centre, ESA, Villafranca del Castillo, Apartado 78, 28691 Villanueva de la Ca\~nada, Spain\\
$^{6}$
Department of Physics, University of Durham, South Road, DH1 3LE, England.
\end{flushleft}

\markboth{
An X-ray Study of Local IR Bright Galaxies
}{
M. Pereira-Santaella et al.
}
\thispagestyle{empty}

\vspace*{0.4cm}
\begin{minipage}[l]{0.09\textwidth}
\ 
\end{minipage}
\begin{minipage}[r]{0.9\textwidth}
\vspace{1cm}
\section*{Abstract}{\small
We are carrying out detailed study of the X-ray and infrared (IR) properties of a sample of local (d $<$ 70\,Mpc) luminous infrared galaxies (LIRGs) using \textit{XMM-Newton} and \textit{Spitzer} (imaging and spectroscopy).
The main goal is to study the extreme processes of star formation and/or active galactic nuclei (AGN) taking place in this cosmologically important class of galaxies.
In this proceedings we present the preliminary results obtained from the analysis of the \textit{XMM-Newton} X-ray images and the X-ray spectral modeling.

\normalsize}
\end{minipage}
\section{Introduction and Sample}

Deep infrared cosmological surveys show that luminous and ultraluminous infrared galaxies (LIRGs, $L_{\rm IR}$ $=$ 10$^{11}$ to 10$^{12}$ $L_{\rm \odot}$, and ULIRGs, $L_{\rm IR}$ $=$ 10$^{12}$ to 10$^{13}$ $L_{\rm \odot}$) are major contributors to the star formation rate density at z\,$\sim$\,1--2 (P\'erez-Gonz\'alez et al. 2005; Le Floc'h et al. 2005; Caputi et al. 2007).
However, unlike local ULIRGs where most of the activity is taking place in very compact ($<$1\,kpc) nuclear regions (Soifer et al. 2001), in z\,$\sim$\,2 infrared bright galaxies the star formation appears to be distributed over spatial scales of a few kpc as in local LIRGs (Daddi et al. 2007; Farrah et al. 2007; Rujopakarn et al. 2010).
Furthermore, to evaluate the different IR methods for selecting AGN in cosmological surveys it is important to understand the X-ray properties of IR-bright galaxies with very high star formation rates (SFR, Donley et al. 2008). LIRGs have SFR in the range 17 to 170\,$M_{\rm \odot}$ yr$^{-1}$ using the prescriptions of Kennicutt (1998). These results highlight the importance of studying the IR and X-ray properties of the population of local LIRGs.
In this conference proceedings we present the first results of an X-ray study of a sample of local LIRGs. The main goals of this project include to relate the X-ray properties to the star-formation activity, and to look for low-luminosity AGN or deeply buried AGN.

The sample consists of 21 local LIRGs with X-ray data drawn from the volume limited sample of local LIRGs (40\,Mpc $<$ d $<$ 75\,Mpc) described in Alonso-Herrero et al. (2006).
We obtained new \textit{XMM-Newton} data for 9 galaxies (proposals 55046, 60160 and 65242) and \textit{XMM-Newton} archive data for the rest. The new observations were for galaxies classified as H\,\begin{small}II\end{small} galaxies based on their optical spectra whereas most of the galaxies from the archive are active galaxies (Seyfert and LINER activity).
These 21 galaxies constitute a representative sample of the LIRG class in terms of both IR luminosity and nuclear activity.

\section{Data Reduction}\label{s:data_reduction}

For the data reduction we followed the standard \textit{XMM-Newton} data reduction procedures. We used SAS version 10.0.2 for the data processing and the latest calibration files (July 2010).
The calibrated event files were generated by the \texttt{emproc} and \texttt{epproc} tasks from the EPIC pn and MOS observations data files (ODF).
Then we filtered out high-background periods to maximize the signal to noise ratio in the 0.5--10\,keV energy range. We used a circular aperture to extract the spectra of the galaxies and we created the redistribution matrices with \texttt{rmfgen}. Finally we rebinned the combined MOS spectrum (MOS1 and MOS2 spectra) and the PN spectrum in order to have at least 15 counts for each spectral bin. Likewise we obtained X-ray images of the galaxies using the EPIC pn calibrated event files. We considered only single and double pixel events and rejected all events which are close to the CCD borders or bad pixels. We adaptively smoothed these images using the SAS task \texttt{asmooth}.

Simultaneously with the X-ray observations we obtained optical and UV images of the galaxies using the \textit{XMM-Newton} optical monitor (OM) with all the available filters (V, B, U, UVW1, UVW2 and UVM2). We used the SAS script \texttt{omichain} for the data reduction. This script processes the OM ODF files and produces calibrated images taking into account the telescope tracking information and the flat fielding corrections.
For some filters there was more than one exposure that we combined to increase the signal to noise ratio.

Due to the low number of counts in the RGS data we could not obtain any spectrum for these galaxies.

\section{X-ray Images and Ultraluminous X-ray Sources}

\begin{figure}[ht]
\center
\includegraphics[width=15.0cm,viewport=11 60 557 224]{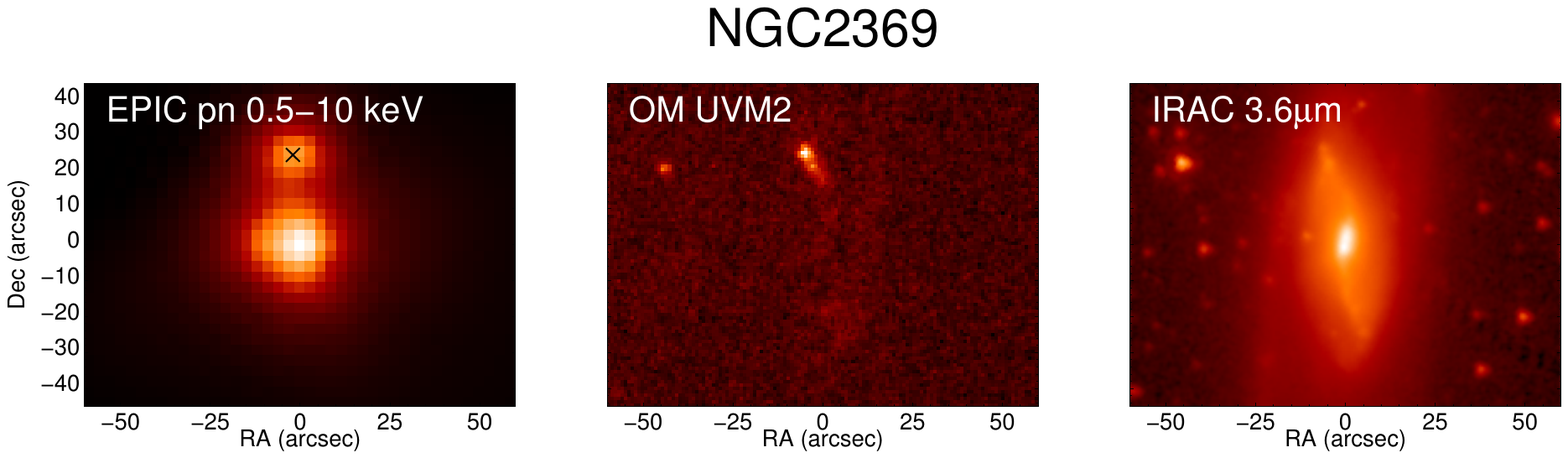}
\includegraphics[width=15.0cm,viewport=11 60 557 224]{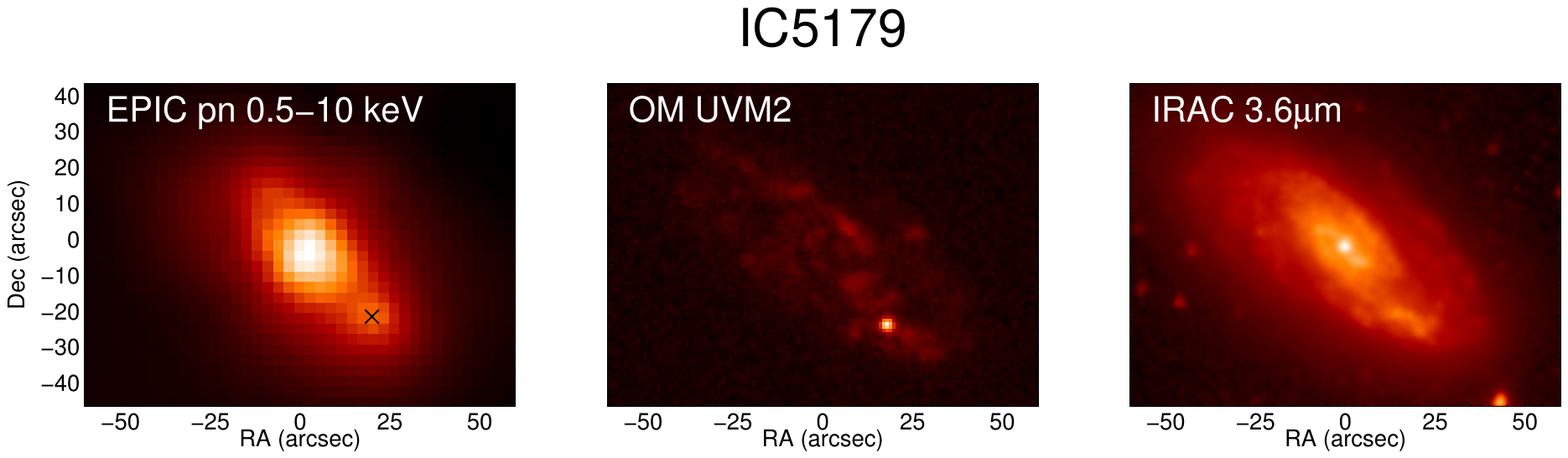}
\caption{\label{fig1}Comparison of the full band X-ray (0.5--10\,keV) \textit{XMM-Newton} EPIC pn (left), UVM2 (2310\,\AA) \textit{XMM-Newton} OM (middle) and 3.6\,$\mu$m \textit{Spitzer}\slash IRAC (right) morphologies for 2 LIRGs in our sample. The black cross in the left panel marks the position of the ULX candidate.
}
\end{figure}

We obtained X-ray images of these LIRGs as described in Section \ref{s:data_reduction}. The left panels of Figure \ref{fig1} show the adaptively smoothed images for 2 of the LIRGs. The middle and right panels show the UV (2310\,\AA) and near-IR \textit{Spitzer}\slash IRAC (3.6\,$\mu$m) images for comparison. For our sample of LIRGs we find different X-ray emission morphologies. Some of them are dominated by the nuclear emission and appear as point like sources at the \textit{XMM-Newton} spatial resolution (5\,arcsec), e.g. NGC~2369 (Figure \ref{fig1} top panels) whereas others are more extended, e.g. IC~5179  (Figure \ref{fig1} bottom panels).
The extended X-ray emission of IC~5179 is accompanied by extended UV emission.

In these two galaxies there are extranuclear X-ray bright sources that might be ultraluminous X-ray sources (ULXs). The ULX candidate in NGC~2369 is located 25\,arcsec ($\sim$5\,kpc) to the north of the nucleus, the candidate in IC~5179 is located 19\,arcsec ($\sim$4\,kpc) to the south-west of the nucleus.
Both sources also appear clearly detected in the UV images (Figure \ref{fig1} middle panels) but not in the 3.6\,$\mu$m images (Figure \ref{fig1} right panels).
We extracted the spectrum of the NGC~2369 ULX candidate. An absorbed power law model reproduces well the observed spectrum ($\chi^2\slash {\rm dof} =1.02$) and implies an intrinsic X-ray luminosity ($L^{int}_{\rm 0.5-8\,keV} = 3.1 \pm 0.4 \times 10^{40}$\,erg s$^{-1}$) one order of magnitude higher than the ULX luminosity threshold ($L^{int}_{\rm 0.5-8\,keV} > 10^{39}$\,erg s$^{-1}$) and comparable to the luminosities measured for other ULXs (Swartz et al. 2004). The parameters of the model ($\Gamma= 1.8 \pm 0.2$ and $N_{\rm H}=1.7 \pm 0.5 \times 10^{21}\,{\rm cm}^{-2}$) are also similar to those obtained for other ULXs (Swartz et al. 2004). We could not extract the spectrum of the ULX candidate in IC~5179 due to the lower number of counts of this source and the underlying extended X-ray emission.

\section{Spectral Modeling}

\begin{figure}[ht]
\center
\includegraphics[width=7.2cm,viewport=39 38 710 580]{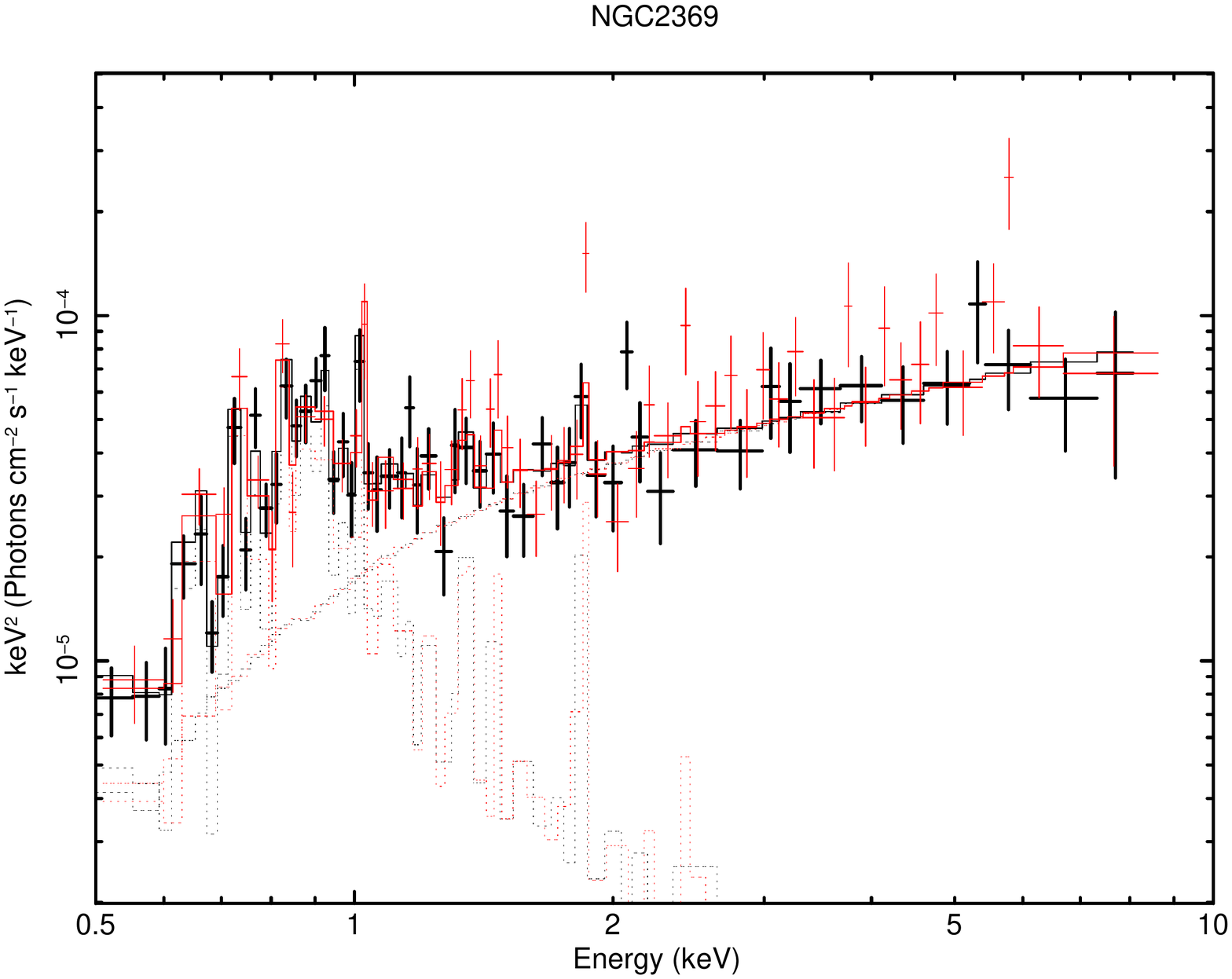}
\includegraphics[width=7.2cm,viewport=39 38 710 580]{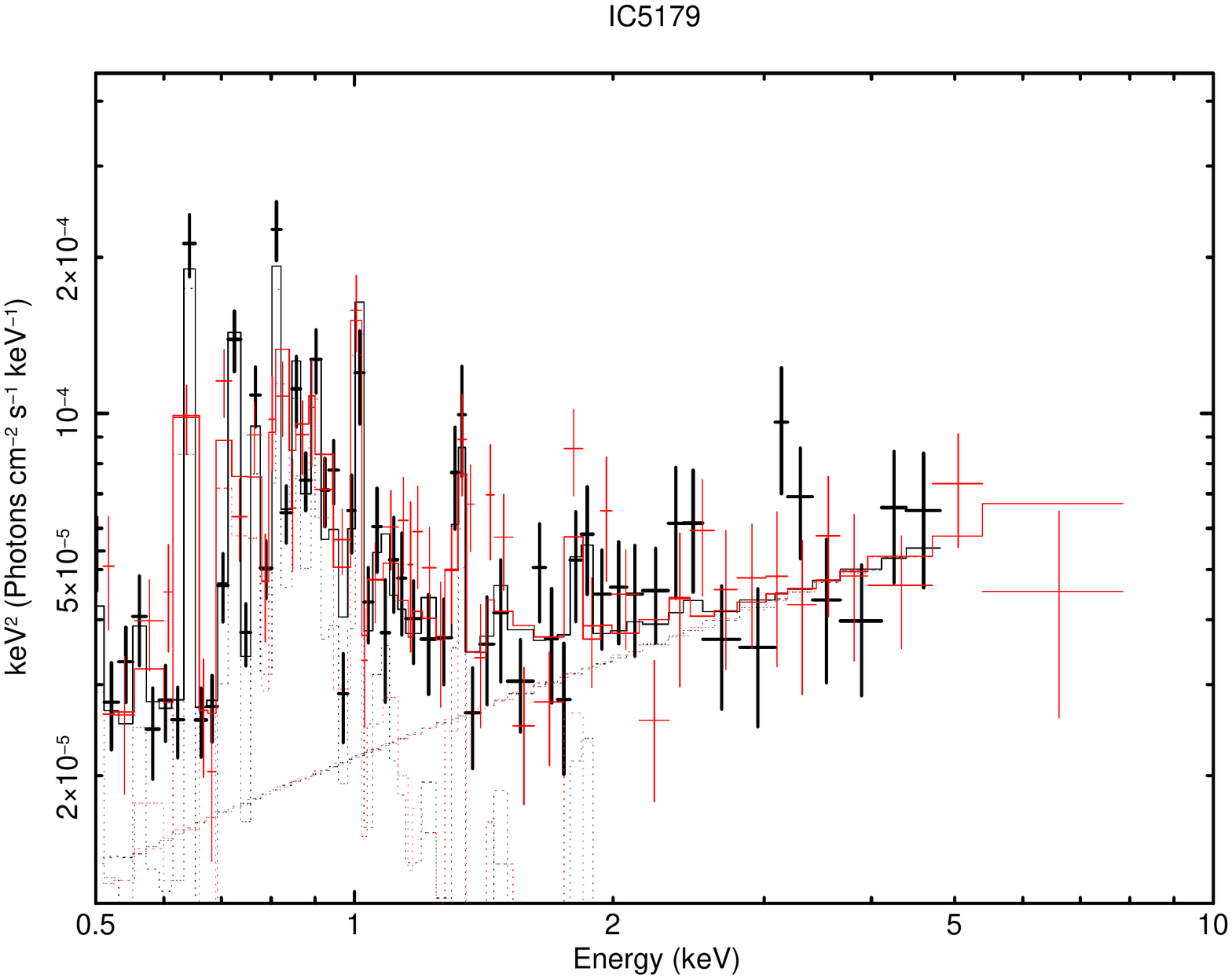}
\caption{\label{fig2} Observed \textit{XMM-Newton} X-ray spectra and best fit models. The black and red lines correspond to the EPIC pn and EPIC MOS data respectively. The model consists of an absorbed power-law plus a soft thermal plasma (\textit{mekal}).
}
\end{figure}

At the typical distance of these LIRGs (d$\sim$60\,Mpc) the \textit{XMM-Newton} spatial resolution (5\,arcsec) corresponds to 1.5\,kpc. This means that we are not able to resolve individual emitting sources. Instead, the \textit{XMM-Newton} nuclear spectra of these LIRGs probably include the emission from X-ray binaries (low- and high-mass binaries), supernova remnants (SNRs), diffuse hot plasma and an AGN may be present as well. Therefore, ideally we would include in the model one component for each one. However this is not possible because: it is complicated to determine the characteristic spectrum of these objects and even more complicated determine the characteristic integrated spectrum of these objects in a galaxy; and the signal to noise ratio of our data is not enough to obtain statistically meaningful results with a very complex model.
For this reasons, we tried to fit the spectra using a simple model consisting of a soft thermal plasma (\textit{mekal}) plus an absorbed power law. This model provides a good fit to the data ($\chi^2$\slash dof $<$ 1.2) for all the galaxies. Figure \ref{fig2} shows the data and the fitted model for 2 of the galaxies.
The plasma abundances were fixed to the solar values except that Fe abundance was left as a free parameter. The typical values of the free model parameters for these galaxies are: $\Gamma$ $\sim$ 1.4--2.2, $N_{\rm H}$ $\sim$ $<$10$^{21}$--10$^{22}$\,cm$^{-2}$, kT $\sim$ 0.4--0.8\,keV and [Fe\slash O] $\sim$ -0.5--0.
The Fe K$\alpha$ emission line is not detected in any of the new galaxies, thus we calculated the upper limits for a narrow emission line ($\sigma$ $=$ 0.1\,keV) at 6.4\,keV.

\small
\section*{Acknowledgments}
This work is based on observations obtained with \textit{XMM-Newton}, an ESA science mission with instruments and contributions directly funded by ESA Member States and the USA (NASA).
MP-S acknowledges support from the CSIC under grant JAE-Predoc-2007. MP-S also thanks the Durham University for their hospitality during his stay while part of this work was done. AA-H and MP-S acknowledge support from the Spanish Plan Nacional del Espacio under grant ESP2007-65475-C02-01.

\end{document}